\documentclass[9pt,twocolumn,twoside]{pnas-new}

\templatetype{pnasresearcharticle} 

\fancypagestyle{firststyle}{
    \fancyhf{}  
}

\begin{document}

\title{Short-range order stabilizes a cubic Fe alloy in Earth's inner core}

\author[a,1]{Zhi Li}
\author[a,1]{Sandro Scandolo}

\affil[a]{The Abdus Salam International Centre for Theoretical Physics, Str. Costiera, 11, Trieste, 34151, Italy}

\leadauthor{Li}


\authordeclaration{The authors declare no competing interest.}
\correspondingauthor{\textsuperscript{1}To whom correspondence should be addressed. E-mail: zli@ictp.it,  scandolo@ictp.it}


\begin{abstract}
The phase diagram and sound velocities of the Fe-Si binary alloy, crucial for understanding the Earth's core, are determined at inner core boundary pressure with \textit{ab-initio} accuracy through deep-learning-aided hybrid Monte Carlo simulations. A complex phase diagram emerges close to the melting temperature, where a re-entrance of the body-centered cubic (bcc) phase is observed. The bcc structure is stabilized by a pronounced short-range ordering of the Si atoms. The miscibility gap between the short-range ordered bcc structure and the long-range ordered cubic B2 structure shrinks with increasing temperature and the transition becomes continuous above 6000 K. We find that a bcc Fe-Si solid solution reproduces crucial geophysical data such as the low shear sound velocity and the seismic anisotropy of the inner core much better than other structures.     
\end{abstract}

\dates{This manuscript was compiled on \today}

\maketitle
\thispagestyle{firststyle}
\ifthenelse{\boolean{shortarticle}}{\ifthenelse{\boolean{singlecolumn}}{\abscontentformatted}{\abscontent}}{}

\firstpage[4]{3}

\dropcap{E}arth's core is composed primarily of iron (Fe) alloyed with small amounts of nickel (Ni) and of light elements such as Si, S, O, C, and H \cite{hirose2021light,hirose2013composition}. Despite their low abundance, light elements play a crucial role in determining the properties of the alloy and in explaining geophysical observations of core density and seismic wave velocities, as well as in the understanding of Earth's accretion and core formation processes \cite{hirose2021light,hirose2013composition}. While the properties of the alloy end-member, pure Fe, at the extreme conditions of pressure and temperature of the Earth's core, are reasonably well understood, 
\cite{kraus2022measuring, tateno2010structure, anzellini2013melting}, phase relations for more realistic compositions for the core are subject to considerable uncertainty even in the case of binary systems, due to intrinsic difficulties in reproducing the relevant conditions of pressure and temperature in the laboratory\cite{hirose2013composition}. Ab-initio atomistic simulations have provided important contributions to our understanding of the Earth's core\cite{alfe1999melting,laio2000physics,vovcadlo2003possible}, but attempts to determine phase relations in core-forming alloys with ab-initio methods are hindered by the large amount of calculations required to sample atomic configurational disorder in solid solutions at high temperature\cite{alfe2002composition,alfe2000constraints}. It has been recently pointed out that even slight departures from a totally random distribution of the elements in the lattice can affect the properties of solid solutions significantly, and that these departures can now be theoretically quantified with ab-initio accuracy \cite{sheriff_2024}.   


The relative concentration of light elements in the Earth's solid inner core is uncertain. However, evidence from the presence of Si in iron meteorites and its depletion in the Earth's mantle suggest that Si could be the most abundant light element in the Earth's core, with concentrations estimated up to 16 atomic percent (at\%), equivalent to 8 weight percent (wt\%) \cite{alfe2002composition, rubie2011heterogeneous, georg2007silicon, allegre1995chemical}. Therefore, crucial insights into the present state and thermodynamic evolution of the core can be gained by constraining the phase diagram and elastic properties of the Fe-Si alloy at Earth's core conditions. 

Experimental studies on the phase relations for the Fe-Si alloy \cite{huang2019equation, wicks2018crystal, ozawa2016high, fischer2013phase, tateno2015structure} remain, with a few exceptions, limited to pressures that are below the range relevant for the Earth's inner core (330-360 GPa). Above 200 GPa, these experiments indicate that the Fe-Si phase diagram comprises the hexagonal close-packed (hcp), body-centered cubic (bcc), B2 (a cubic phase isostructural to CsCl), and liquid phases. However, the precise location of the phase boundaries at core pressures remains elusive, and as a consequence, attempts to infer the chemical composition and crystal structure of the Earth's inner core from the phase diagram are fraught with significant ambiguities. 
A heated debate persists whether it exists as a two-phase mixture (hcp+B2) or it comprises solely the bcc or hcp phase \cite{tateno2015structure, fischer2013phase, ozawa2016high}. The solubility limit of Si in solid Fe and the size of the miscibility gap between end-members (Fe and FeSi) are also poorly constrained, especially close to Earth's core temperatures and pressures. At lower pressures (40 GPa), experiments indicate a strong temperature dependence of the phase boundaries and a stabilization of the bcc phase at temperatures close to melting temperatures \cite{edmund2022fe}. The distinction between the two reported cubic structures (bcc and B2) is also poorly characterized. 
The B2 crystal symmetry is a sub-group of the bcc symmetry caused by the emergence of a distinction between the two simple-cubic sublattices that compose the bcc structure. Evidence for a discontinuous transition between two distinct structures \cite{dubrovinsky2003,fischer2013phase} was not confirmed by more recent experiments \cite{lin2009phase,tateno2015structure}. Although the B2 and bcc phases are structurally similar, it was recently pointed out that while the sound velocity of the bcc phase is consistent with seismic values\cite{Li_GRL}, the B2 phase has significantly higher values that are incompatible with geophysical observations\cite{nagaya2023equations}. 

Theoretical investigations of the phase diagram of solid Fe-Si alloys using first-principle methods \cite{cote2008light, cote2010ab, vovcadlo2003possible, Belonoshko2009, cui2013effect} have yielded conflicting results regarding the crystal structure of the alloy. Earlier work \cite{alfe2000constraints,alfe2002composition} suggests that the structure of the alloy is the same (hcp) as that of pure Fe. However, in later studies the bcc \cite{vovcadlo2003possible, Belonoshko2009} and face-centered cubic (fcc) structures \cite{cote2010ab} have been reported to be stabilized by small concentrations of Si over the hcp structure. No theoretical attempt has been made to clarify structural differences and relative stability of the B2 and bcc phases at core conditions. 

While first-principle accuracy is required to discriminate between different structures, accurately sampling vibrational and configurational disorder in a solid solution is challenged by the large size and number of the physical realizations of the system that need to be calculated in order to achieve statistical converged values of free energies. As a result, most previous first-principle studies have implicitly assumed a random distribution of the Fe and Si atoms, primarily based on the observation that Fe and Si possess very similar atomic sizes at high pressure\cite{alfe2002composition}. However, this idealized assumption has been recently challenged by the direct experimental observations of chemical short-range order (SRO) in high-entropy alloys composed of elements with comparable atomic sizes \cite{chen2021direct, zhang2020short}. The existence of SRO violates the ideal-mixing approximation and can have a significant impact on the thermodynamic properties and phase relations of a solid solution \cite{alam2011phonon, gao2017thermodynamics}. In the specific case of Fe-Si, Alfe et al.\cite{alfe2000constraints,alfe2002composition} have included the effects of non-ideal mixing, but only in the limit of small Si concentrations. However, none of the theoretical work reported so far has been able to include the full effects of configurational disorder in the solid solution at arbitrary concentrations, with first-principle accuracy. 

We recently developed a deep-learning-based method to accelerate the sampling of both the vibrational and configurational disorder in alloys, while retaining first-principles accuracy in the description of the interatomic potential \cite{Li_CPC}. The method combines particle swaps performed with Monte Carlo (MC) moves with deterministic molecular dynamics (MD) simulations to sample the thermal agitation of particles within a given lattice configuration. It enables the determination of Gibbs free energy surfaces of solid solutions with density-functional theory (DFT) accuracy\cite{kohn1965self,hohenberg1964inhomogeneous,mermin1965thermal}. Here we apply this method to construct phase relations in the Fe-Si binary alloy over a large range of concentrations up to 1:1 stoichiometry, at Earth's core conditions of pressure and temperature.   


\begin{figure*}[htbp]
    \centering
    \includegraphics[width=0.7\linewidth]{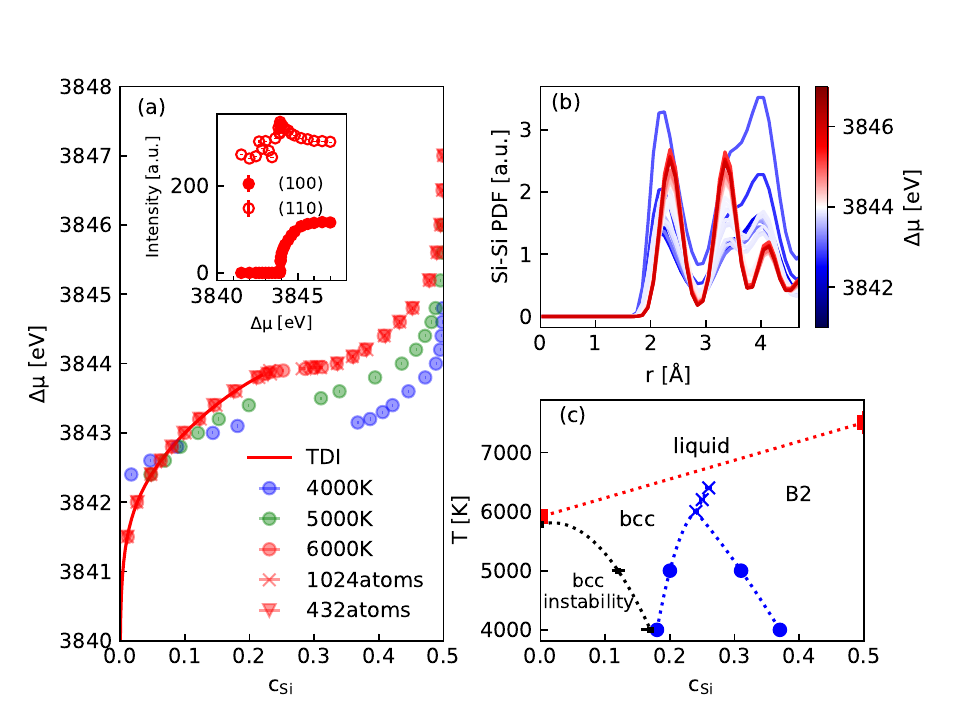}
    \caption{Hybrid Monte-Carlo simulations of cubic Fe$_{1-x}$Si$_x$ solid solutions as a function of the chemical potential difference between Si and Fe ($\Delta \mu$). Panel (a) shows the calculated Si concentration as a function of $\Delta \mu$ at different temperatures. The inset displays the simulated X-Ray diffraction  intensity of the (100) and (110) peaks at 6000 K, the (100) peak being a fingerprint of the B2 phase. 
    Convergence on system size was checked by running simulations  with 432 and 1024 atoms. Panel (b) shows the evolution of the Si-Si pair distribution function as function of $\Delta \mu$. Panel (c) presents a schematic diagram of the Fe-Si system restricted to the bcc, B2 and liquid phases. The black dashed line shows the temperature below which we find the bcc structure to be mechanically unstable. 
    } 
    \label{fig:semi}
\end{figure*}

\subsection*{Bcc-B2 phase transition}
Prompted by previous theoretical studies suggesting that the presence of Si can stabilize the bcc structure over hcp\cite{Belonoshko2009}, we begin our study of the Fe-Si solid solutions by restricting our analysis to cubic lattices, and investigating order-disorder transitions in the bcc and B2 structures. FeSi, as a 1:1 stochiometric compound, crystallizes in the B2 structure at high pressures, in which the two simple-cubic sublattices of the parent bcc structure are fully occupied by Fe and Si atoms, respectively. When the Si concentration is intermediate between the end-members Fe and FeSi, the location of the Si atoms is determined by a balance between the competing effects of entropy, which favors a fully disordered state where the probability of finding a Si atom in the two sublattices is identical, and interatomic energies, which favor an ordered, FeSi-like, state where Si occupies preferentially one of the two sublattices of the bcc structure. The group-subgroup symmetry relation between the bcc and B2 structures gives rise to extra peaks ([100], [210], etc) in the x-ray diffraction pattern of the B2 phase, with respect to the set of reflections that characterize the bcc structure ([110], [200], [220], etc)\cite{lin2009phase,fischer2013phase}. 

We simulate the transition between bcc and B2 in the Fe-Si system using the semi-grand canonical (SGC) ensemble, where the total number of particles and the difference in chemical potential, denoted as $\Delta \mu = \mu_{\mathrm{Fe}} - \mu_{\mathrm{Si}}$, remain fixed, while the relative concentration of the two species is allowed to fluctuate by swapping the identity of an atom from Fe to Si, or vice versa. 
The simulations were conducted using a 3456-atom simulation box at temperatures of 4000 K, 5000 K, and 6000 K, respectively, and at a pressure of 330 GPa, with the initial structure being a disordered bcc. By varying the values of $\Delta \mu$, the equilibrium Si concentration was extracted. 

As illustrated in Fig.~\ref{fig:semi}(a), we observe a discontinuity in the Si concentration at $\Delta \mu =$ 3843.1 and 3843.4 eV for temperatures of 4000 K and 5000 K, respectively. This indicates the presence of a first-order phase transition from the bcc to the B2 structure. At low concentrations the probability of Si atoms occupying one of the two simple-cubic sublattices is the same, consistent with a disordered bcc structure. At higher concentrations Si atoms occupy preferentially one of the sublattices, giving rise to a structure with B2 crystal symmetry and to the appearance of extra peaks in the X-ray diffraction (XRD) spectra. At 6000 K, the change in Si concentration becomes continuous. In the inset of Fig.~\ref{fig:semi}(a) we show the XRD intensities of the [100] and [110] peaks across the transition at 6000 K. The [100] peak characteristic of the B2 structure appears at about $\mu \sim$ 3844 eV, corresponding to a Si concentration of 24 at\%. Changes in the long-range ordering of the structure reflect also changes in the short-range atomic structure of the solid solution, as shown in Fig.~\ref{fig:semi}(b). The first peak of the Si-Si pair distribution function includes the 1st and 2nd coordination shell of the bcc structure. Its shift to higher distances with increasing $\Delta\mu$ reflects the depletion of the 1st shell arising from the B2-like ordering of the Si sublattice. The relative height of the third and fourth peaks, located at 3.4 and 4.2 $\mathrm{\AA}$, respectively, is also consistent with the bcc-to-B2 transition. Fig.~\ref{fig:semi}(c) shows a sketch of the Fe-Si phase diagram restricted to the bcc-B2 system, based on our simulations. The miscibility gap shrinks with increasing temperature and the observed change in the order of the phase transition from discontinuous to continuous between 5000 K and 6000 K hints at the presence of a tricritical point occurring below 6000 K, above which the transition becomes second order. The phenomena observed in our study align well with previous work on the bcc Fe-Al system using a model Hamiltonian \cite{Dunweg_Binder}. 

We also found that at high temperatures, the B2-type ordering is present even at concentrations as low as 16 wt\% Si, which significantly deviates from stoichiometric FeSi. Furthermore, our analysis at 6000 K highlights that although Si atoms in the B2 structure preferentially occupy one of the two cubic sublattices, some degree of disorder due to sublattice chemical interchange persists. 

In our recent work on pure Fe ($c_{\mathrm{Si}}=0$) \cite{Li_GRL} we have shown that range of temperatures where bcc Fe is mechanically stable is small. Bcc Fe becomes mechanically unstable only a few hundred degreees Kelvin below its melting point, due to the vanishing of the $c_{11}-c_{12}$ shear elastic constant. We computed the elastic constants of the Fe-Si solid solution and find that increasing concentrations of Si lead to a dramatic lowering of the temperature at which the bcc structure becomes unstable and therefore to a substantially broadening of the temperature range where the bcc structure is mechanically stable (see Fig.~\ref{fig:semi}(c)).   

\subsection*{Fe-Si phase diagram at 330 GPa}
Having clarified the nature of the order-disorder transition in the bcc-B2 system, we are now in a position to extend our study to include all other structures (hcp, fcc, liquid) and determine the full phase diagram of the Fe-Si alloy. To this aim, we employ our recently developed HMC-based thermodynamic integration (TDI) method to compute the free energies of the hcp, fcc, and liquid phases with first-principle accuracy. 

\begin{figure*}
    \centering
    \includegraphics[width=0.8\linewidth]{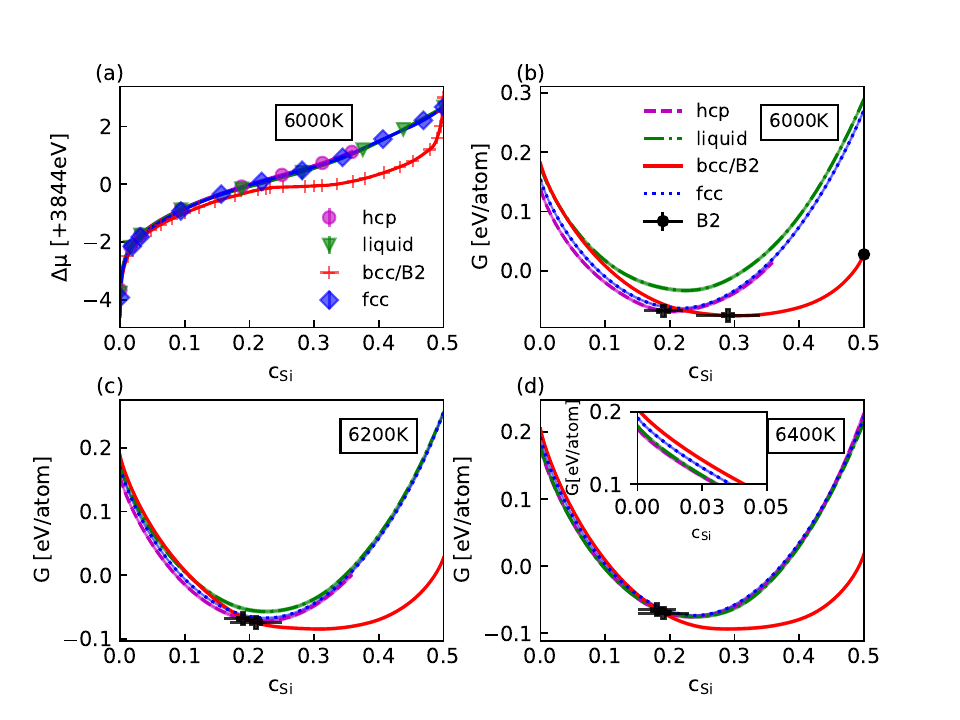}
    \caption{Calculated Gibbs free energies for Fe$_{1-x}$Si$_x$ solid solution with the bcc, B2, fcc, hcp structure, as well for the liquid phase, at different temperatures. (a) Gradient of the free energy with respect to c$_{Si}$ obtained from the first-stage TDI simulations. (b)-(d) Gibbs free energies at different temperatures. The black crosses represent the cotangent points between bcc/B2 and hcp phases.}
    \label{fig:free_energy_calculation_6000K}
\end{figure*}

We show the results of the HMC-TDI simulations at 6000 K and 330 GPa in Fig.~\ref{fig:free_energy_calculation_6000K}(a). These simulations were conducted at several Si concentrations, and a smooth curve was fitted to the resulting $\Delta \mu$, following the approach proposed by \citep{kranendonk1991free,alfe2002composition}. As $\Delta \mu$ represents the derivative with respect to $c_{\textrm{Si}}$ of the free energy difference between two phases connected by the transmutation of one Fe atom into a Si atom, its integration over $c_{\textrm{Si}}$ yields the free energy difference between the solutions and the pure phase, which is shown in Fig.~\ref{fig:free_energy_calculation_6000K}(b). The continuous nature of the bcc-B2 phase transition can also be inferred from the observation that starting from bcc Fe and integrating along $\Delta \mu$ - $\mathrm{c_{Si}}$ up to 0.5 results in a free energy that differs from the B2 phase from direct TDI simulations by only 2 meV/atom. Additionally, a characteristic flat region near the minimum in the Gibbs free energy curve is indicative of a continuous phase transition.

Phase boundaries are determined by identifying the common tangent line between two relevant phases. At 6000 K (Fig.~\ref{fig:free_energy_calculation_6000K}(b)), the solid hcp phase remains stable up to 21 at\% Si, beyond which a miscibility gap emerges. For Si concentrations exceeding 30 at\%, only the B2 phase is stable. As the phase transition boundary between bcc and B2 occurs at approximately 24 at\% Si, the bcc phase is not present at this temperature. Moreover, we observe that the fcc phase closely aligns with the free energy of hcp iron at high concentration of Si, but remains less stable than the B2 phases. Additionally, the Gibbs free energy curve of hcp iron closely parallels but remains lower than that of the liquid phase, suggesting the nearly congruent melting of the hcp alloy with a partition coefficient close to 1, in agreement with the results of Alfe et al.\cite{alfe2002composition}.
We applied the same strategy to determine the phase boundary at temperatures of 4000, 5000, 6000, 6200, and 6400 K, respectively. 

\begin{figure*}
    \centering
    \includegraphics[width=0.7\linewidth]{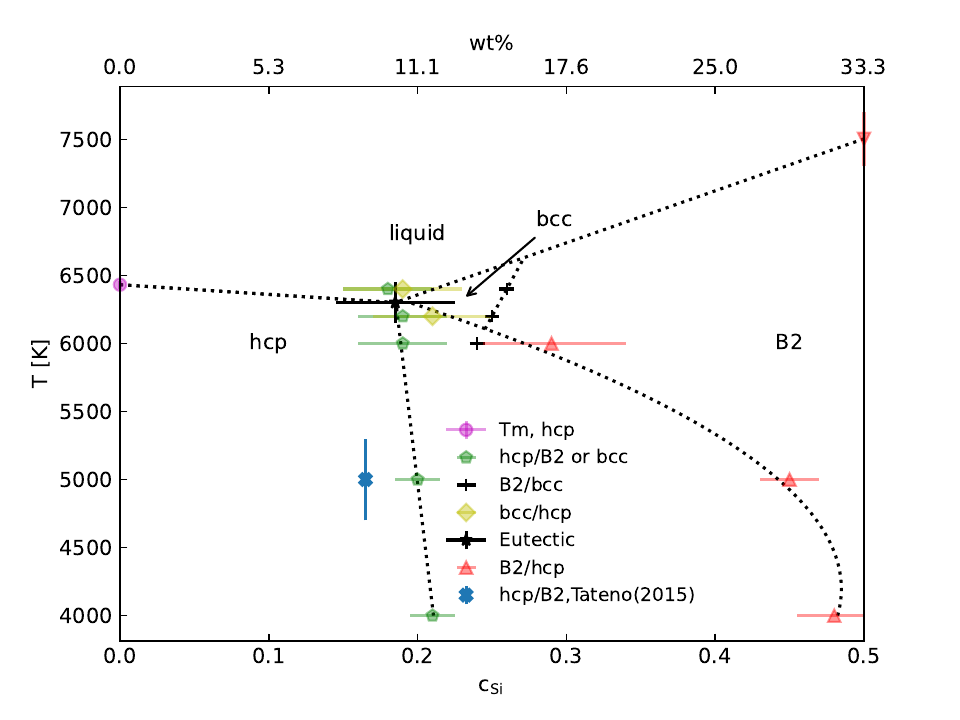}
    \caption{Phase diagram of the Fe-Si binary alloy at the Earth's inner core boundary pressure of 330 GPa and compared with experimental studies\cite{tateno2015structure}. The eutectic point is located at a temperature of 6300 $\pm$ 150 K and a Si concentration of 10 $\pm$ 1.5 wt\%. 
    }
    \label{fig:phase_diagram}
\end{figure*}

The full phase diagram is illustrated in Fig.~\ref{fig:phase_diagram}. Below 6000 K, a miscibility gap between the hcp and B2 phases is evident, which is consistent with the only available experimental data at this pressure\cite{tateno2015structure}. As temperature increases, the phase relationships become more complex due to the strong temperature dependence of the phase boundary of the cubic phases and to the appearance of the bcc phase above 6200 K, which pushes the stability region of the hcp phase towards lower $c_{\textrm{Si}}$ and the B2 phase towards higher $c_{\text{Si}}$ at high temperatures. The boundaries of the stability range of the bcc structure are subject to a large uncertainty, but clearly show that at Earth's inner core temperatures the bcc structure is stable in the range 20-25 Si at\%, much lower than the concentration required to stabilize the cubic phase at lower temperatures. We determine the eutectic point to be at 19 $\pm$ 3 at\% and 6300 $\pm$ 150 K. The eutectic Si concentration is in the middle of the hcp/bcc miscibility gap, while the eutectic temperature should lie somewhere between 6200 K, where hcp iron alloy remains more stable than the liquid phase (Fig.~\ref{fig:free_energy_calculation_6000K}(c)) and 6450 K, where pure hcp iron melts. The finding that the eutectic temperature of the alloy is close to the melting temperature of pure Fe implies a solid-liquid partition coefficient of Si close to one, and suggests that Si alone cannot explain the observed density jump across the inner-core boundary. Moreover, in the assumption that the Earth's inner core is composed of a single phase since its formation, our study puts an upper limit of 10 wt\% for the maximum Si content in the Earth's liquid outer core (out of which the inner core crystallizes). 

\subsection*{The effects of short-range ordering on phase stability}
\begin{figure}
    \centering
    \includegraphics[width=1.0\linewidth]{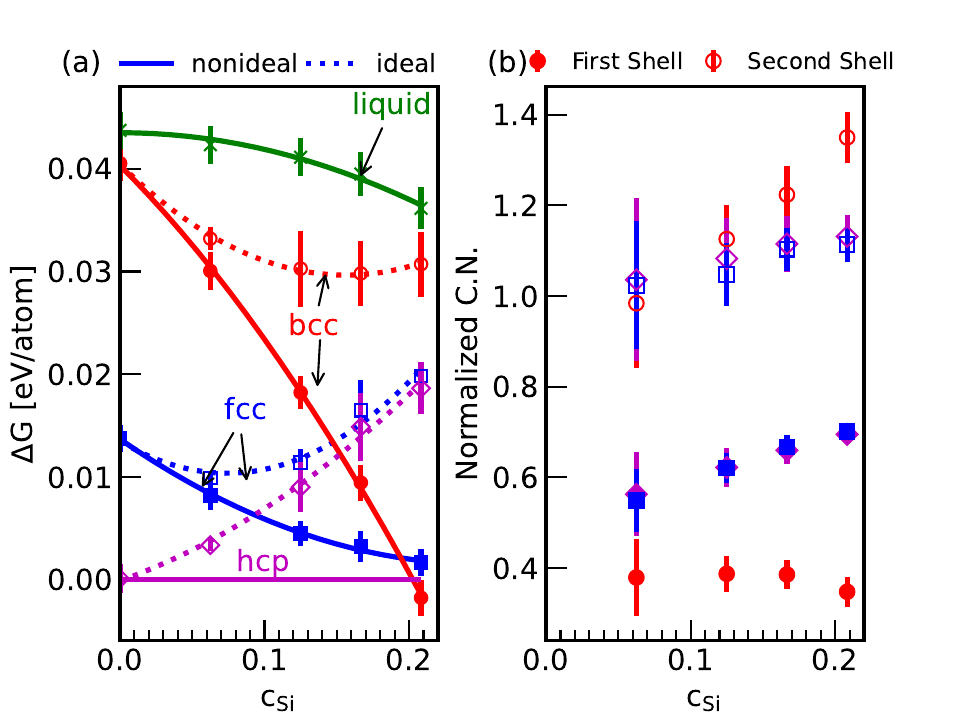}
    \caption{The effects of short-ranged ordering on the free energy for the solid Fe-Si phases are shown. (a) displays the Gibbs free energy relative to the non-ideal hcp phase. The free energy of ideal solid solutions was determined through the TDI method, while the free energy of non-ideal solid solutions was determined using the HMC-TDI method. For the latter case, the identity interchange Monte Carlo move allows sampling of the full configurational space. The error bars for the ideal solid solutions represent the standard deviation of the obtained free energy for five different configurations by randomly assigning Si in the simulation box. (b) shows the normalized coordination number (C.N.) for the Si-Si pair in the first and second coordinate shell with respect to a fully disordered random alloy.}
    \label{fig:SRO}
\end{figure}

While the presence of a solid solutions with B2 symmetry at $c_{\text{Si}}\lesssim 0.5$ and hcp symmetry at $c_{\text{Si}}\gtrsim 0$ is not surprising in view of the structures of the end-members, the presence in the window around 6200-6500 K and 20-25 at\% Si of a stability domain of the bcc structure, that is of a lattice where the location of the Si atoms looses the B2 long-range order, indicates that other, more complex  types of ordering of the Si sublattice must come into play in order to explain the reversal of the hcp-bcc relative stability when the Si concentration increases from $c_{\text{Si}} =  0$ to $c_{\text{Si}} \sim 0.2$. 

Most previous theoretical works have assumed that the location of the Si atoms in the bcc lattice is purely random. To understand the extent to which the ''ideal mixing'' approximation holds, we determined the free energy of solid phases by forcing a random distribution of the Si atoms in the bcc lattice, and calculating free energies using the standard thermodynamic integration method. 
The comparison between between the ideal-mixing approximation and the fully self-consistent non-ideal calculation  is shown in Fig.~\ref{fig:SRO}. As expected, the non-ideal solid solutions have a lower free energy compared with the ideal ones, and their difference increases with increasing Si concentration. For Fe-10 wt\% Si, the free energy difference for hcp Fe-Si can be as large as 19 meV/atom, and it becomes even larger for the bcc phase, reaching about 34 meV/atom. At $c_{\text{Si}} =0.2$ the effect is large enough to compensate the free-energy difference between hcp and bcc in pure Fe, and for $c_{\text{Si}} > 0.2$  the bcc solid solution becomes more stable than the hcp solid solution. As shown in Fig.\ref{fig:SRO}(b), the significant difference between non-ideal and ideal solid solutions results from the presence of SRO in the relative position of the Si atoms, which tends to decrease the probability of its presence in the first coordination shell and increase it in the second one. We found that the effects of SRO are much stronger for the bcc phase than for the hcp and fcc phases, and stabilize the bcc phase at Earth's inner conditions. 


\subsection*{A cubic Fe alloy in the Earth's inner core?}
\begin{figure}
    \centering
    \includegraphics[width=\linewidth]{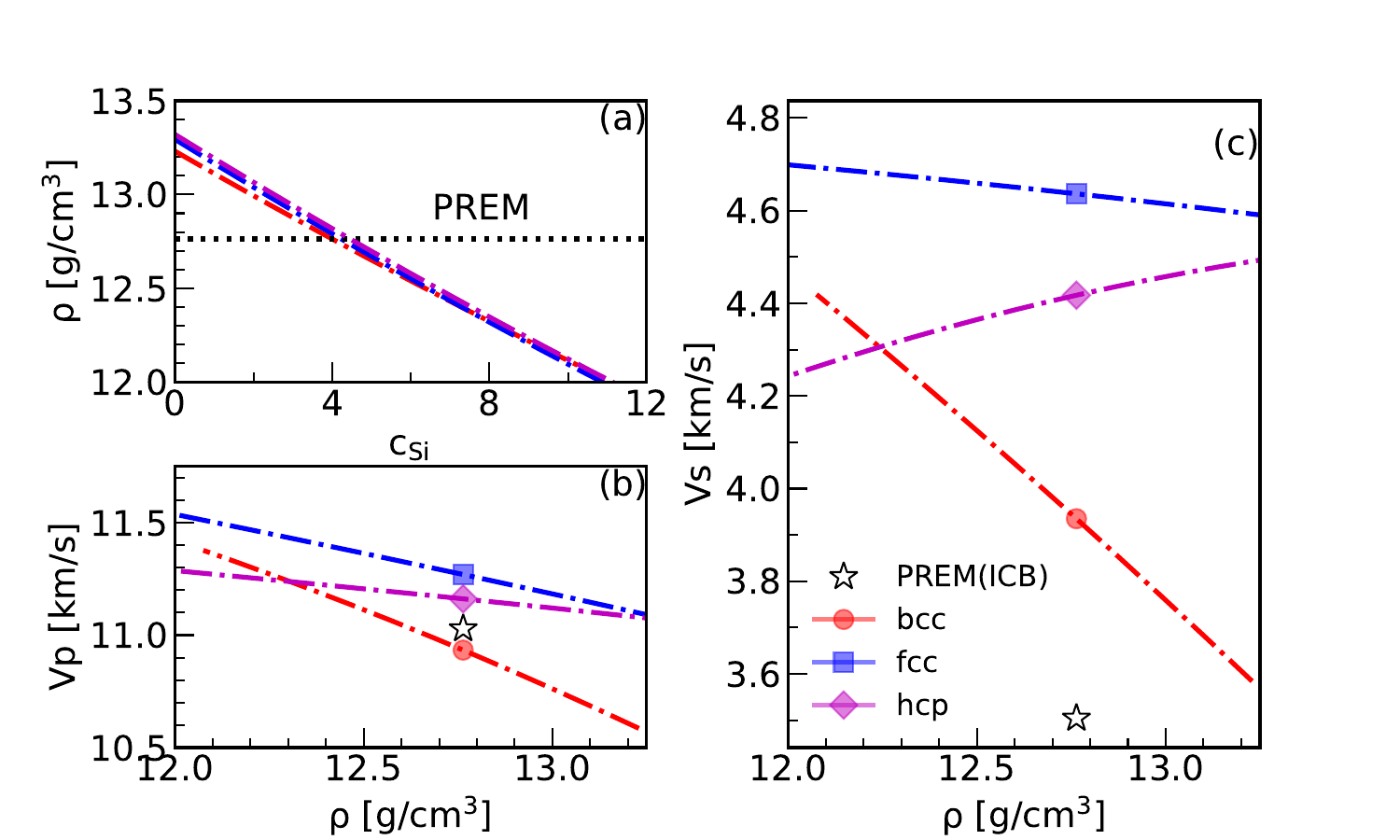}
    \caption{The effects of Si on the density (a), compressional wave velocity (b), and shear wave velocity (c) for the bcc, fcc, and hcp iron alloys at 6000 K and 330 GPa and compared to the Preliminary Earth Reference Model (PREM) \cite{dziewonski1981preliminary}.}
    \label{fig:sound_velocity}
\end{figure}

A direct comparison of our results with geophysical observations for the inner core \cite{dziewonski1981preliminary} is obtained by computing densities and sound velocities of all the solid Fe-Si phases considered in this work, at the pressure and temperature conditions of the inner-core boundary, 330 GPa and 6000 K (Fig.~\ref{fig:sound_velocity}). We find that Si has a similar effect on the density of different structures, causing a decrease of $\sim 0.1$ g/cm$^3$ per 1 wt\% Si. The Si concentration required to match the density of the Earth's inner core at 330 GPa is 5 wt\%, independent of the crystal structure. For hcp, the result agrees with the findings of Alfe et al.\cite{alfe2002composition}. This concentration is well within the stability rage of the hcp structure (Fig.~\ref{fig:phase_diagram}), a result which is in contrast with the mixture of B2 and hcp phases proposed in previous studies for similar Si concentrations but lower pressures \cite{ikuta2021two}.

The effect of Si on compressional wave velocities is small for all solid phases, with only a slight increase of less than 4\% as Si concentration increases from 0 to 10 wt\%. We conclude that the comparison of density and compressional wave velocities with geophysical data is not sufficient to unequivocally discriminate between different crystal structures for the core-forming alloy. In contrast, we find that different solid structures show greater relative variations in shear wave velocities. The best agreement with the PREM model \cite{dziewonski1981preliminary} is obtained for the bcc alloy, whereas the shear velocity of the hcp (fcc) alloy is 25\% (30\%) higher than seismic data. It has been recently proposed that a superionic Fe:H hcp structure could yield values for the shear velocity that are consistent with PREM\cite{he2022superionic}. However, the calculated anisotropy of the compressional wave velocities in superionic hcp (about 6\%) is too close to the observed anisotropy of seismic compressional waves (waves propagating along the Earth's polar direction are 3–4\% faster than those in the equatorial plane), which would would imply a large degree of crystal orientation of superionic hcp in the inner core. By contrast, the compressional wave anisotropy of the bcc alloy is about 20\%, fully compatible with seismic observations\cite{mattesini2010hemispherical}. 

Our finding that short-range order provides an additional, unexpected mechanism for the stabilization of a bcc solid solution, and that a bcc alloy matches geophysical data for density and seismic velocities better than any other structure, suggests that the possibility that the Earth's inner core adopts a bcc structure cannot be ruled out, especially considering that the effects of Ni and other light elements were not included in this study. 
As shown in Fig.~\ref{fig:semi}(c), at high temperature the bcc phase can dissolve up more than 10 wt\% Si without transitioning into the B2 phase. This covers the entire range of possible Si concentrations for the Earth's core \cite{georg2007silicon}. 
With about Fe-5 wt\% Si, the free energy difference between a bcc and an hcp solid solution at 6000 K is only 23 meV/atom (see Fig.~\ref{fig:SRO}(a)), an amount that could be easily overturned by small concentrations of other elements in the solid solution.  
Indeed, recent ab initio simulations and experimental studies have revealed a strong stabilizing effect of Ni on the bcc phase \cite{sun2024unveiling,ikuta2021two}. The additional effects of C, H, and O could also be significant, though their concentrations in the Earth's inner core remain uncertain. Future studies are imperative to extend the deep-learning model to even more realistic geophysically scenarios, including ternary systems and beyond. Despite the computational challenges, determining these phase diagrams of multi-component systems with our approach is feasible and could provide valuable insights into the composition and structure of the Earth's inner core.

\matmethods{
\subsection*{The Construction and validation of the deep-learning interatomic potentials}
We employed an active-learning method similar to our previous work \cite{li2024deep} to construct deep-learning interatomic potentials (DLP) for the B2, bcc, fcc, hcp, and liquid Fe-Si alloys, with the training dataset built iteratively. Energies, forces, and virial stress tensors for these structures were computed using Quantum ESPRESSO \cite{giannozzi2009quantum, giannozzi2017advanced}. The atom identity swap Monte-Carlo move was also employed to expedite the sampling of the configurational space. As shown in Li and Scandolo \cite{li2024deep}, temperature-dependent thermal electron excitations significantly affect the stability of solid phases. To accurately replicate the DFT potential energy surface, we constructed a total of five DLP models at temperatures of 4000, 5000, 6000, 6200, and 6400 K. To reduce computational workload, we initially constructed the DP model from scratch at 6000 K and then refined the model for the other temperatures. Across all conditions, the root-mean-squared errors of energy, pressure, and force were below 8.5 meV/atom, 0.7 GPa, and 0.38 eV/\AA, respectively. Detailed information on the active-learning process, the number of structures in the training dataset, neural network architecture, and DFT simulation parameters can be found in Text S1 and S2 of the Supporting Information (SI).

We used an independent testing dataset to validate the developed DLP models and found that their performance remains consistent across all Si concentrations and phases, and comparable to that on the training dataset (See Table S1 in SI). Additionally, we conducted validation checks with up to 500 atoms for the fcc, hcp, and liquid phases, and up to 1024 atoms for the bcc phase, confirming that the tail of the interatomic potential was accurately captured, suggesting that our DLP model can be applied to large-scale simulations (See Text S3 in SI). Furthermore, we performed accuracy checks using free energy perturbation theory and found that the discrepancy between the DLP model and DFT potentials was less than 13 meV/atom (See Fig. S3 in SI). We also verified the performance of the DLP models on elastic properties, which showed good agreement with DFT data (See Text S4 in SI). Therefore, the trained DLP models are accurate proxies for the DFT potentials. 

\subsection*{Hybrid Monte Carlo method}
We employ the hybrid Monte Carlo method coupled with deep-learning interatomic potentials \citep{duane1987hybrid, mehlig1992hybrid} to sample the phase space, allowing to treat positional and configurational disorders on an equal footing. Depending on the targeted statistical ensemble, we use various Monte Carlo moves. For the canonical ensemble, we apply atom displacement MC moves and identity interchange MC sweeps. In the atom displacement MC move, a short molecular dynamics (MD) simulation of $N_{\text{md}}=5$ steps in the microcanonical ensemble is used to generate the trial configuration. In the identity interchange MC sweep, up to 5\% of the atoms undergo identity interchange moves, where each move involves randomly selecting one Fe atom and one Si atom and swapping their positions. Note that in an identity interchange MC sweep, only the final configurations contribute to the ensemble average. For the isobaric-isothermal ensemble, we use a composite MC displacement move that combines atom displacement and volume fluctuation, in addition to the identity interchange MC sweeps. The trial configuration in the composite displacement move is generated by running an MD simulation in the isoenthalpic–isobaric ensemble for $N_{\text{md}}=5$ steps, based on the algorithm proposed in \cite{martyna1994constant}. All Monte Carlo moves are accepted or rejected based on the Metropolis criterion \cite{metropolis1953equation}. Detailed technical aspects can be found in our recent work \cite{Li_CPC}. In practice, we define two probabilities, $\eta_{\mathrm{disp}}=0.75$ and $\eta_{\mathrm{identity}}=0.25$, to determine the choice among the different moves. To quantify the uncertainty in the reported values, we employ the block averaging method \cite{Flyvbjerg1989} and use the Monte Carlo method for error propagation \cite{ANDERSON19761533}.

\subsection*{Semi-grand isobaric and isothermal ensemble}
To understand the order-disorder phenomenon in the bcc-FeSi system, we simulate the solid solutions using the semi-grand isobaric and isothermal ensemble ($\Delta \mu PT$) as developed by \cite{kofke1988}, where $P$ and $T$ represent pressure and temperature, respectively. In this approach, the total number of particles and the chemical potential difference $\Delta \mu = \mu_{\text{Fe}} - \mu_{\text{Si}}$ are kept fixed while allowing the compositions to fluctuate. This method is advantageous compared to the conventional canonical and isobaric-isothermal ensembles, where two phases can coexist over a wide range of compositions. In contrast, at a fixed $\Delta \mu$, only one phase remains stable (bcc vs. B2) in the $\Delta \mu PT$ ensemble. The simulation procedure is similar to the hybrid Monte Carlo algorithm in the $NPT$ ensemble. Instead of the atom identity MC sweep in the semi-grand ensemble, the atom transmutation MC sweep is applied, which consists many atom transmutation MC moves up to 5 \% of the total number of atoms. We define two probabilities, $\eta_{\mathrm{disp}}=0.75$ and $\eta_{\mathrm{transmutation}}=0.25$, for the choice among the different moves. The MC transmutation of atomic species is accepted by the Metropolis criterion\cite{metropolis1953equation}, 
\begin{equation}
  p = \min[1, \exp(-\beta(\Delta U - \Delta \mu \Delta N_{Si}))],  
\end{equation}
where $\Delta N_{Si} = 1$ if a Fe atom is transmuted into Si, while $\Delta N_{Si} = -1$ if a Si atom is transmuted into Fe, $\beta = {(k_BT)}^{-1}$ where $k_B$ is the Boltzmann constant, and $\Delta U$ is the change in potential energy evaluated using the deep-learning interatomic potential. 

\subsection*{Thermodynamic integration for solutions}
To determine the phase diagram for the Fe-Si system, we employ a four-step procedure to compute the free energies of the bcc, B2, fcc, hcp, and liquid phases by transforming a system with an analytic free energy to the one described by a DFT potential. The phase space is sampled using the hybrid Monte Carlo method, ensuring no approximation is made on the configurational entropy of the solid phases \cite{Li_CPC}. Except for the bcc or B2 phases, in the first step, we construct a reversible thermodynamic path connecting the solutions to their pure phase counterparts and compute their free energy difference ($\Delta G_1 = G_{\text{solution}}^{\text{DLP}} - G_{\text{pure}}^{\text{DLP}}$) by using the DLP models. For the bcc or B2 phase, the relationship between $\Delta \mu$ and Si concentration ($c_{\mathrm{Si}}$) can be obtained from simulations in the semi-grand canonical ensemble. Since $\Delta \mu$ represents the free energy gradient, $\Delta G_1$ can be extracted by integrating the obtained $\Delta \mu$ as a function of $c_{\mathrm{Si}}$. We have also verified the consistency between the thermodynamic integration and the semi-grand canonical ensemble methods, and found excellent agreement as expected (Fig.~\ref{fig:semi}).

In the second step, we apply the thermodynamic integration method to determine the Gibbs free energy difference between the DLP model and a modified version of the Lennard-Jones potential that has a soft repulsive core \cite{beutler1994avoiding} ($\Delta G_2 = G_{\text{pure}}^{\text{DLP}} - G_{\text{pure}}^{\text{LJ}}$). In the third step, we use the Frenkel-Ladd method \cite{frenkel1984new} to compute the free energy difference between the solid or liquid pure phase and a reference system ($\Delta G_3 = G_{\text{pure}}^{\text{LJ}} - G_{\text{pure}}^{\text{ref}}$). For solid and liquid iron, the reference systems are chosen as the Einstein crystal and the ideal gas, respectively. In the final step, we calculate the free energy difference described by a DLP model compared to DFT based on free energy perturbation theory ($\Delta G_4 = G_{\text{solution}}^{\text{DFT}} - G_{\text{solution}}^{\text{DLP}}$), ensuring that the obtained free energy fully reaches \textit{ab initio} accuracy (see Text S3 in SI). Consequently, the free energy of a solution is the sum of all the free energy differences calculated in these four steps. We have previously benchmarked this method for the MgO-CaO system under ambient pressure conditions, and it agrees very well with available experimental results \cite{Li_CPC}.

The finite size effects in the first two steps are mitigated by using large simulation cells with 3456, 3072, 4000, and 3456 atoms for bcc/B2, hcp, fcc, and liquid iron alloys, respectively. In the third step, systematic finite-size scaling is performed to obtain the free energy values for the LJ potential in the thermodynamic limit (see, e.g., Figure 2 in \cite{Li_CPC}). In the final step, we used simulation cells with 432, 500, 448, and 432 atoms for the bcc/B2, fcc, hcp, and liquid $\mathrm{Fe_{1-x}Si_x}$ phases, respectively. Simulations were performed for an interval of $x=0.0625$, and data were interpolated using a piecewise cubic polynomial to obtain $\Delta G_4$ at other $x$ values. Additional tests with a 1024-atom cell for bcc iron yielded results consistent with those obtained from the 432-atom simulation cell.

\subsection*{Elastic properties}
The elastic constants of hcp, fcc, and bcc iron alloys at 6000 K and 330 GPa were calculated using the stress-strain method, as described in our previous study \cite{Li_GRL}. In this method, a small strain is applied to a well-equilibrated simulation cell, and the resulting deviatoric stress is determined by performing hybrid Monte Carlo simulations in the canonical ensemble. As demonstrated in our previous study \cite{Li_GRL}, simulation cells with 3456, 4000, and 3072 atoms for bcc, fcc, and hcp, respectively, yield converged values for the elastic constants.

}
\showmatmethods{} 

\acknow{We acknowledge CINECA (Consorzio Interuniversitario per il Calcolo Automatico) under the ISCRA initiative and the Leonardo early access program, for the computing resources and support. LZ also thanks Chao Li (ICTP) for many useful discussions on statistical sampling with the hybrid Monte Carlo method.}

\showacknow{} 


\bibliography{pnas-sample}

\begin{thebibliography}{10}

\bibitem{hirose2021light}
K Hirose, B Wood, L Vo{\v{c}}adlo, Light elements in the earth’s core.
\newblock {\em\protect\JournalTitle{Nature Reviews Earth \& Environment}} \textbf{2}, 645--658 (2021).

\bibitem{hirose2013composition}
K Hirose, S Labrosse, J Hernlund, Composition and state of the core.
\newblock {\em\protect\JournalTitle{Annual Review of Earth and Planetary Sciences}} \textbf{41}, 657--691 (2013).

\bibitem{kraus2022measuring}
RG Kraus, et~al., Measuring the melting curve of iron at super-earth core conditions.
\newblock {\em\protect\JournalTitle{Science}} \textbf{375}, 202--205 (2022).

\bibitem{tateno2010structure}
S Tateno, K Hirose, Y Ohishi, Y Tatsumi, The structure of iron in earth’s inner core.
\newblock {\em\protect\JournalTitle{Science}} \textbf{330}, 359--361 (2010).

\bibitem{anzellini2013melting}
S Anzellini, A Dewaele, M Mezouar, P Loubeyre, G Morard, Melting of iron at earth’s inner core boundary based on fast x-ray diffraction.
\newblock {\em\protect\JournalTitle{Science}} \textbf{340}, 464--466 (2013).

\bibitem{alfe1999melting}
D Alfe, M Gillan, G Price, The melting curve of iron at the pressures of the earth's core from ab initio calculations.
\newblock {\em\protect\JournalTitle{Nature}} \textbf{401}, 462--464 (1999).

\bibitem{laio2000physics}
A Laio, S Bernard, G Chiarotti, S Scandolo, E Tosatti, Physics of iron at earth's core conditions.
\newblock {\em\protect\JournalTitle{Science}} \textbf{287}, 1027--1030 (2000).

\bibitem{vovcadlo2003possible}
L Vo{\v{c}}adlo, et~al., Possible thermal and chemical stabilization of body-centred-cubic iron in the earth's core.
\newblock {\em\protect\JournalTitle{Nature}} \textbf{424}, 536--539 (2003).

\bibitem{alfe2002composition}
D Alf{\`e}, M Gillan, GD Price, Composition and temperature of the earth’s core constrained by combining ab initio calculations and seismic data.
\newblock {\em\protect\JournalTitle{Earth and Planetary Science Letters}} \textbf{195}, 91--98 (2002).

\bibitem{alfe2000constraints}
D Alfe, M Gillan, G Price, Constraints on the composition of the earth's core from ab initio calculations.
\newblock {\em\protect\JournalTitle{Nature}} \textbf{405}, 172--175 (2000).

\bibitem{sheriff_2024}
K Sheriff, Y Cao, T Smidt, R Freitas, Quantifying chemical short-range order in metallic alloys.
\newblock {\em\protect\JournalTitle{Proceedings of the National Academy of Sciences}} \textbf{121}, e2322962121 (2024).

\bibitem{rubie2011heterogeneous}
DC Rubie, et~al., Heterogeneous accretion, composition and core--mantle differentiation of the earth.
\newblock {\em\protect\JournalTitle{Earth and Planetary Science Letters}} \textbf{301}, 31--42 (2011).

\bibitem{georg2007silicon}
RB Georg, AN Halliday, EA Schauble, BC Reynolds, Silicon in the earth’s core.
\newblock {\em\protect\JournalTitle{Nature}} \textbf{447}, 1102--1106 (2007).

\bibitem{allegre1995chemical}
CJ Allegre, JP Poirier, E Humler, AW Hofmann, The chemical composition of the earth.
\newblock {\em\protect\JournalTitle{Earth and Planetary Science Letters}} \textbf{134}, 515--526 (1995).

\bibitem{huang2019equation}
H Huang, et~al., Equation of state for shocked fe-8.6 wt\% si up to 240 gpa and 4,670 k.
\newblock {\em\protect\JournalTitle{Journal of Geophysical Research: Solid Earth}} \textbf{124}, 8300--8312 (2019).

\bibitem{wicks2018crystal}
JK Wicks, et~al., Crystal structure and equation of state of fe-si alloys at super-earth core conditions.
\newblock {\em\protect\JournalTitle{Science advances}} \textbf{4}, eaao5864 (2018).

\bibitem{ozawa2016high}
H Ozawa, K Hirose, K Yonemitsu, Y Ohishi, High-pressure melting experiments on fe--si alloys and implications for silicon as a light element in the core.
\newblock {\em\protect\JournalTitle{Earth and Planetary Science Letters}} \textbf{456}, 47--54 (2016).

\bibitem{fischer2013phase}
RA Fischer, et~al., Phase relations in the fe--fesi system at high pressures and temperatures.
\newblock {\em\protect\JournalTitle{Earth and Planetary Science Letters}} \textbf{373}, 54--64 (2013).

\bibitem{tateno2015structure}
S Tateno, Y Kuwayama, K Hirose, Y Ohishi, The structure of fe--si alloy in earth's inner core.
\newblock {\em\protect\JournalTitle{Earth and Planetary Science Letters}} \textbf{418}, 11--19 (2015).

\bibitem{edmund2022fe}
E Edmund, et~al., The fe-fesi phase diagram at mercury’s core conditions.
\newblock {\em\protect\JournalTitle{Nature Communications}} \textbf{13}, 387 (2022).

\bibitem{dubrovinsky2003}
L Dubrovinsky, et~al., Iron--silica interaction at extreme conditions and the electrically conducting layer at the base of earth's mantle.
\newblock {\em\protect\JournalTitle{Nature}} \textbf{422}, 58--61 (2003).

\bibitem{lin2009phase}
JF Lin, et~al., Phase relations of fe-si alloy in earth's core.
\newblock {\em\protect\JournalTitle{Geophysical Research Letters}} \textbf{36} (2009).

\bibitem{Li_GRL}
Z Li, S Scandolo, Competing phases of iron at earth's core conditions from deep-learning-aided \textit{ab-initio} simulations (2024) Geophysical Research Letters(Accepted).

\bibitem{nagaya2023equations}
Y Nagaya, H Gomi, K Ohta, K Hirose, Equations of state for b2 and bcc fe1-xsix.
\newblock {\em\protect\JournalTitle{Physics of the Earth and Planetary Interiors}} \textbf{341}, 107046 (2023).

\bibitem{cote2008light}
AS C{\^o}t{\'e}, L Vo{\v{c}}adlo, JP Brodholt, Light elements in the core: Effects of impurities on the phase diagram of iron.
\newblock {\em\protect\JournalTitle{Geophysical research letters}} \textbf{35} (2008).

\bibitem{cote2010ab}
AS C{\^o}t{\'e}, L Vo{\v{c}}adlo, DP Dobson, D Alf{\`e}, JP Brodholt, Ab initio lattice dynamics calculations on the combined effect of temperature and silicon on the stability of different iron phases in the earth's inner core.
\newblock {\em\protect\JournalTitle{Physics of the Earth and Planetary Interiors}} \textbf{178}, 2--7 (2010).

\bibitem{Belonoshko2009}
AB Belonoshko, A Rosengren, L Burakovsky, DL Preston, B Johansson, Melting of fe and ${\text{fe}}_{0.9375}{\text{si}}_{0.0625}$ at earth's core pressures studied using ab initio molecular dynamics.
\newblock {\em\protect\JournalTitle{Phys. Rev. B}} \textbf{79}, 220102 (2009).

\bibitem{cui2013effect}
H Cui, Z Zhang, Y Zhang, The effect of si and s on the stability of bcc iron with respect to tetragonal strain at the earth's inner core conditions.
\newblock {\em\protect\JournalTitle{Geophysical Research Letters}} \textbf{40}, 2958--2962 (2013).

\bibitem{chen2021direct}
X Chen, et~al., Direct observation of chemical short-range order in a medium-entropy alloy.
\newblock {\em\protect\JournalTitle{Nature}} \textbf{592}, 712--716 (2021).

\bibitem{zhang2020short}
R Zhang, et~al., Short-range order and its impact on the crconi medium-entropy alloy.
\newblock {\em\protect\JournalTitle{Nature}} \textbf{581}, 283--287 (2020).

\bibitem{alam2011phonon}
A Alam, RK Chouhan, A Mookerjee, Phonon modes and vibrational entropy of disordered alloys with short-range order: A first-principles calculation.
\newblock {\em\protect\JournalTitle{Physical Review B}} \textbf{83}, 054201 (2011).

\bibitem{gao2017thermodynamics}
MC Gao, et~al., Thermodynamics of concentrated solid solution alloys.
\newblock {\em\protect\JournalTitle{Current Opinion in Solid State and Materials Science}} \textbf{21}, 238--251 (2017).

\bibitem{Li_CPC}
Z Li, S Scandolo, Efficient determination of free energies of non-ideal solid solutions via hybrid monte carlo simulations.
\newblock {\em\protect\JournalTitle{Computer Physics Communications}} \textbf{304}, 109307 (2024).

\bibitem{kohn1965self}
W Kohn, LJ Sham, Self-consistent equations including exchange and correlation effects.
\newblock {\em\protect\JournalTitle{Physical review}} \textbf{140}, A1133 (1965).

\bibitem{hohenberg1964inhomogeneous}
P Hohenberg, W Kohn, Inhomogeneous electron gas.
\newblock {\em\protect\JournalTitle{Physical review}} \textbf{136}, B864 (1964).

\bibitem{mermin1965thermal}
ND Mermin, Thermal properties of the inhomogeneous electron gas.
\newblock {\em\protect\JournalTitle{Physical Review}} \textbf{137}, A1441 (1965).

\bibitem{Dunweg_Binder}
B D\"unweg, K Binder, Model calculations of phase diagrams of magnetic alloys on the body-centered-cubic lattice.
\newblock {\em\protect\JournalTitle{Phys. Rev. B}} \textbf{36}, 6935--6952 (1987).

\bibitem{kranendonk1991free}
W Kranendonk, D Frenkel, Free energy calculations for solid solutions by computer simulations.
\newblock {\em\protect\JournalTitle{Molecular physics}} \textbf{72}, 699--713 (1991).

\bibitem{dziewonski1981preliminary}
AM Dziewonski, DL Anderson, Preliminary reference earth model.
\newblock {\em\protect\JournalTitle{Physics of the earth and planetary interiors}} \textbf{25}, 297--356 (1981).

\bibitem{ikuta2021two}
D Ikuta, E Ohtani, N Hirao, Two-phase mixture of iron--nickel--silicon alloys in the earth’s inner core.
\newblock {\em\protect\JournalTitle{Communications Earth \& Environment}} \textbf{2}, 225 (2021).

\bibitem{he2022superionic}
Y He, et~al., Superionic iron alloys and their seismic velocities in earth’s inner core.
\newblock {\em\protect\JournalTitle{Nature}} \textbf{602}, 258--262 (2022).

\bibitem{mattesini2010hemispherical}
M Mattesini, et~al., Hemispherical anisotropic patterns of the earth’s inner core.
\newblock {\em\protect\JournalTitle{Proceedings of the National Academy of Sciences}} \textbf{107}, 9507--9512 (2010).

\bibitem{sun2024unveiling}
Y Sun, et~al., Unveiling the effect of ni on the formation and structure of earth’s inner core.
\newblock {\em\protect\JournalTitle{Proceedings of the National Academy of Sciences}} \textbf{121}, e2316477121 (2024).

\bibitem{li2024deep}
Z Li, S Scandolo, Deep-learning interatomic potential for iron at extreme conditions.
\newblock {\em\protect\JournalTitle{Physical Review B}} \textbf{109}, 184108 (2024).

\bibitem{giannozzi2009quantum}
P Giannozzi, et~al., Quantum espresso: a modular and open-source software project for quantum simulations of materials.
\newblock {\em\protect\JournalTitle{Journal of physics: Condensed matter}} \textbf{21}, 395502 (2009).

\bibitem{giannozzi2017advanced}
P Giannozzi, et~al., Advanced capabilities for materials modelling with quantum espresso.
\newblock {\em\protect\JournalTitle{Journal of physics: Condensed matter}} \textbf{29}, 465901 (2017).

\bibitem{duane1987hybrid}
S Duane, AD Kennedy, BJ Pendleton, D Roweth, Hybrid monte carlo.
\newblock {\em\protect\JournalTitle{Physics letters B}} \textbf{195}, 216--222 (1987).

\bibitem{mehlig1992hybrid}
B Mehlig, D Heermann, B Forrest, Hybrid monte carlo method for condensed-matter systems.
\newblock {\em\protect\JournalTitle{Physical Review B}} \textbf{45}, 679 (1992).

\bibitem{martyna1994constant}
GJ Martyna, DJ Tobias, ML Klein, Constant pressure molecular dynamics algorithms.
\newblock {\em\protect\JournalTitle{The Journal of chemical physics}} \textbf{101}, 4177--4189 (1994).

\bibitem{metropolis1953equation}
N Metropolis, AW Rosenbluth, MN Rosenbluth, AH Teller, E Teller, Equation of state calculations by fast computing machines.
\newblock {\em\protect\JournalTitle{The journal of chemical physics}} \textbf{21}, 1087--1092 (1953).

\bibitem{Flyvbjerg1989}
H Flyvbjerg, HG Petersen, {Error estimates on averages of correlated data}.
\newblock {\em\protect\JournalTitle{The Journal of Chemical Physics}} \textbf{91}, 461--466 (1989).

\bibitem{ANDERSON19761533}
G Anderson, Error propagation by the monte carlo method in geochemical calculations.
\newblock {\em\protect\JournalTitle{Geochimica et Cosmochimica Acta}} \textbf{40}, 1533--1538 (1976).

\bibitem{kofke1988}
DA Kofke, ED Glandt, Monte carlo simulation of multicomponent equilibria in a semigrand canonical ensemble.
\newblock {\em\protect\JournalTitle{Molecular Physics}} \textbf{64}, 1105--1131 (1988).

\bibitem{beutler1994avoiding}
TC Beutler, AE Mark, RC van Schaik, PR Gerber, WF Van~Gunsteren, Avoiding singularities and numerical instabilities in free energy calculations based on molecular simulations.
\newblock {\em\protect\JournalTitle{Chemical physics letters}} \textbf{222}, 529--539 (1994).

\bibitem{frenkel1984new}
D Frenkel, AJ Ladd, New monte carlo method to compute the free energy of arbitrary solids. application to the fcc and hcp phases of hard spheres.
\newblock {\em\protect\JournalTitle{The Journal of chemical physics}} \textbf{81}, 3188--3193 (1984).

\end{thebibliography}

\end{document}